\documentstyle[aps,12pt,epsfig,tighten]{revtex}

\newcommand{\be}{\begin{equation}}
\newcommand{\ee}{\end{equation}}
\newcommand{\bea}{\begin{eqnarray}}
\newcommand{\eea}{\end{eqnarray}}
\newcommand{\ba}{\beta}
\newcommand{\n}{\nu}
\newcommand{\vvr}{\vec{r}}
\newcommand{\vvv}{\vec{v}}
\newcommand{\cn}{{\cal{N}}}
\newcommand{\half}{\frac 1 2}
\begin{document}

\title{A Note on the Ruelle Pressure for a Dilute
Disordered Sinai Billiard}
\author{Henk van Beijeren} 
\address{Institute for Theoretical Physics\\
Utrecht University\\ 
Leuvenlaan 4, 3584 CE Utrecht, The Netherlands\\
and \\
Centre d' Etudes des Ph\'enom\`enes Non-lin\'eaires et des 
Syst\`emes Complexes\\ 
Universit\'e Libre de Bruxelles\\ 
Campus de la Plaine, Boulevard du Triomphe, 1050 Bruxelles, Belgium}
\author{J.\ R.\ Dorfman}
\address{Institute for Physical Science and Technology 
and Department of Physics\\ 
University of Maryland\\
College Park, MD 20742, USA}

\date{\today}

\maketitle

\begin{abstract}
The topological pressure is evaluated for a dilute random
Lorentz gas, 
in the approximation that takes into 
account only uncorrelated collisions between
the moving particle and fixed, hard sphere scatterers. The
pressure 
is obtained analytically as a function of the
temperature-like parameter, $\beta$, and of the density of scatterers,
$n$. The effects of correlated collisions on the topological pressure can be 
described
qualitatively, at least, and they significantly modify the results
obtained by considering only uncorrelated collision sequences.
As a consequence, for large systems, the range of $\beta$-values over which our expressions for the topological pressure are valid becomes very small, 
approaching zero,
in most cases, 
as the inverse of the logarithm of system size.
\end{abstract}

\section{Introduction}
One of the most important developments in dynamical systems theory
over the past several decades is the discovery by Sinai, Ruelle, and
Bowen, that Gibbs ensembles have as central a role to play in the
description of hyperbolic dynamical systems, as they do in statistical
thermodynamics~\cite{sinai,ruelle,bowen}.
The dynamical analog of the Gibbs formalism for
statistical thermodynamics is usually referred to as the {\em thermodynamic
formalism}~\cite{ruelle,gas}.
One of the central quantities in the application of
Gibbs ensembles to dynamical systems is the so-called topological
pressure, or Ruelle pressure, which 
is the analog, apart from a sign, 
of
the Helmholtz free energy per particle of a thermodynamic system, in
the thermodynamic limit. This
correspondence is made clear by its definition in terms of a dynamical
partition function, where the pressure is defined in essentially the
same way as the free energy per particle, with time $t$ playing the
role of 
system size.

In the dynamical case, the analog of the
thermodynamic limit is the infinite time limit
\footnote{In addition one may consider the limit of infinite system size for the 
topological pressure as well. As we will see this leads to new complications 
which do 
not occur in the Gibbsian formalism of equilibrium statistical mechanics}. 
Like the free energy,
as a function of temperature, volume and particle number, generates the full 
equilibrium 
thermodynamics of a many particle system,
the 
topological pressure and its derivatives with respect to a temperature like
parameter, provide useful information about the dynamical properties of the
system under study. Kolmogorov-Sinai and topological entropies per
unit time can be expressed in terms of 
this pressure and its
derivatives, as can escape rates, and Hausdorff dimensions of fractal
structures and measures~\cite{gas,bs}. 

Despite the beauty of the theory and the power of the theorems that
have been proven about the dynamical Gibbs ensembles, there are few
examples of 
deterministic 
dynamical systems where the topological pressure can be evaluated 
by
analytical methods, other than simple maps, such as baker maps and
toral automorphisms~\cite{kathas}.
The present authors, together with Appert and Ernst
calculated the topological pressure for a more complicated dynamical
system, a Lorentz lattice gas~\cite{ap1,ap2}.
Here a moving particle hops 
from 
from one site
to a neighboring site on a lattice, at each time step. Some of the lattice sites 
are 
occupied by scatterers, such
that if the particle encounters a scatterer, it may change the
direction of its motion at the next step, with some given
probability.
If the particle lands on a site that does not have a
scatterer, it continues moving in the same direction at the next time
step. While this model can be 
reformulated as a deterministic dynamical
system, it was possible to evaluate the topological pressure by
treating the system as a stochastic one, and using the 
thermodynamic formalism as it applies to stochastic systems~\cite{gas}.
In the large system limit the topological pressure turns out to be
determined by
either large
dense clusters of scatterers
(for the inverse temperature like parameter 
$\beta<1$),
or large vacant regions 
(for $\beta>1$), that form in a 
quenched distribution of
scatterers on the lattice.
This phenomenon is 
very analogous to the
one responsible for {\em Lifshitz tails} in 
disordered systems showing critical phenomena\cite{lifs,griffs,dv}.

The purpose of this note is to 
provide a similar analysis of the
topological pressure for a very useful model for studies in the kinetic
theory of gases, the Lorentz gas. This model served as the inspiration
for the Lorentz lattice gas described above, but is a continuum
model. It is constructed by placing fixed scatterers in $d$ spatial
dimensions, and allowing a particle to move in this system of
scatterers, colliding with them, and moving freely between collisions.
When the scatterers are placed on a regular lattice this model is known as the 
Sinai billiard~\cite{sinaibil}, which has served as a paradigm in the theory of 
dynamical 
systems for several decades now. It is one of the few nontrivial dynamical 
systems
that are really close to realistic physical systems and yet allow
for rigorous proofs
of many important properties, including the existence of a finite diffusion 
coefficient\cite{chernov}.
For the application of kinetic theory methods, however, it is simpler to   
consider 
disordered systems, where the scatterers are located at random positions
in space.
Further simplifications occur when the density of scatterers is low, so that the 
average 
distance
between them
is much larger than their radii.
In this case the average time for return of the moving particle to a scatterer 
with which it has previously  
collided becomes very long and therefore the effects of correlations 
between 
collisions, to a first approximation, may be ignored in many applications.
The moving
particle makes specular elastic collisions with the scatterers, and
travels freely between collisions. 
At low density of scatterers, in large but not too large systems (we will become 
more 
specific about this in section~\ref{sec:largesys}) with periodic boundary 
conditions 
(the simplest choice; in fact the precise nature of the boundaries does not 
matter so 
much, as long as they are elastic) it is very reasonable to assume that 
subsequent 
collisions of the light particle may be considered as independent random   
flights, 
sampled from the equilibrium distribution of random flights in the dilute 
d-dimensional 
Lorentz gas. This will be the basis for our calculations in 
section~\ref{sec:dpf}, where
we will describe the model in more detail and
define the dynamical partition function and the topological
pressure. Then we show that the contributions from the 
supposedly uncorrelated collision sequences can be
accounted for in a relatively simple way, in the spirit of Sinai's
analysis of billiard ball models. The following section shows how the
topological pressure can then be evaluated in terms of the 
poles of a
zeta function, called the Ruelle zeta function. We
discuss the
properties of 
this pressure and evaluate the relevant dynamical
properties of the Lorentz gas in this approximation.

For large systems it is to be expected that the topological
pressure will be determined by large clusters of scatterers or large
vacant regions, as in the 
case of the Lorentz lattice gas. This will be investigated in 
section~\ref{sec:largesys}, where we will make some specific estimates for the 
topological pressure in large systems both for $\beta>1$ and for $\beta<1$. 
These are 
based on plausible assumptions about the types of closed orbits (for $\beta>1$) 
or dense 
regions (for $\beta<1$) that will dominate under these conditions.
The paper
concludes with a discussion of 
similar large system properties to be expected in systems of many moving   
particles, with 
some remarks on possible relations
between the topological pressure for 
$\beta>1$ and 
scars in the corresponding quantum mechanical models.

\section{The Dynamical Partition Function for the Random Lorentz 
Gas}\label{sec:dpf}
We consider an infinitely extended system of fixed hard-sphere
scatterers of radius $a$, placed at random with number density $n$ and
without overlapping, in a space of $d$
dimensions. A point particle, of mass $m$, moves with speed $v$ in this array of
scatterers making specular, elastic collisions with the scatterers,
and traveling freely between collisions. This system, the random
Lorentz gas, is therefore a
Hamiltonian system, with the usual symplectic properties. We consider here the 
case of 
low density, specified by the
condition that the radius of a scatterer is small compared to the
average distance between two scatterers, {\it i.\
e.} $na^d <<1$. For
the
two and three dimensional version of this model
the Lyapunov exponents and Kolmogorov-Sinai
(KS) entropies
have been calculated, with the use of
kinetic theory methods, appropriate for low
densities of the scatterers~\cite{2d,3d}. Here we apply the same
methods for the calculation of the topological pressure.

The topological pressure is given in terms of a dynamical partition
function, $Z(\ba,t)$, defined, as a function of 
an inverse-temperature-like
parameter $\ba$ and time $t$, by 
\be
Z(\ba,t) = \int d\,\mu(\vvr,\vvv)[\Lambda(\vvr,\vvv,t)]^{(1-\ba)}.
\label{1}
\ee
Here the integration is over the equilibrium measure, $\mu(\vvr,\vvv)$, for the 
phase
space of the moving particle. The position and velocity of the
particle are denoted by $\vvr,\vvv$, respectively, where $\vvr$ ranges over the 
configuration space region available to the moving particle, not
occupied by scatterers, and $\vvv$ ranges over all possible
directions of the velocity, while the magnitude of the velocity
remains constant. The quantity $\Lambda(\vvr,\vvv,t)$ is the
``stretching'' factor for a phase space trajectory of the particle
starting at $\vvr,\vvv$ and 
extending over a time $t$. The stretching
factor is the factor by which the projection of an infinitesimal phase space
volume onto the unstable directions will expand over a time $t$. For
{\em very long times} the stretching factor is given in terms of local positive
Lyapunov exponents, $\lambda_i(\vvr,\vvv)$ by
\be
\Lambda(\vvr,\vvv,t) \approx e^{t\sum_{i}\lambda_i
(\vvr,\vvv)},
\label{2}
\ee
where the subscript $i$ labels the distinct unstable manifolds in
phase space for the moving particle. The 
number of these unstable manifolds is
given in terms of the spatial dimension as $d-1$, since there are
two neutral directions in phase space, and 
equal numbers of unstable
and unstable manifolds. Finally, $\ba$
can be used to determine various dynamical properties
of the system. 
The topological pressure,
or Ruelle pressure, $P(\ba)$ is expressed in terms of the
dynamical partition function by
\be
P(\ba) = \lim_{t\rightarrow\infty}\frac{1}{t} \ln Z(\ba,t),
\label{3}
\ee
which establishes the formal connection between the topological
pressure in dynamics and the negative of the Helmholtz free energy in
statistical thermodynamics. 

The structure of the dynamical partition function, given by
Eq. (\ref{1}) suggests the procedure for its calculation for a Lorentz
gas. One classifies the regions of phase space according to the number
of collisions the moving particle will suffer in time $t$ starting
from a point in that region. Thus we would need to calculate the
contributions to $Z(\ba,t)$ from regions with no collisions, one
collision, and so on. For each such region we then need to calculate
the associated stretching factors.
These stretching factors have been given for the dilute, random Lorentz
gas by 
Van Beijeren, Latz, and Dorfman, as products of stretching
factors, $\Lambda^{(d)}_i$ for each collision of the moving particle. 
They depend on the dimension, $d$, of the system and are easily obtained,
for low densities (long free flight times), by methods described in~\cite{2d,3d}
, ass
\be
\Lambda_i^{(d)}(\tau_i,\theta_i) =
\left[\frac{2v\tau_i}{a}\right]^{(d-1)}|\cos\theta_{i}|^{(d-3)}
\label{4}
\ee
Here $\tau_i$ is the free time between the collision denoted by the
subscript $i$, and the previous collision of the moving particle, and
$\theta_i$ is the angle of incidence  at collision $i$. It will be
convenient here to express this angle in terms of the inner product of the
incident velocity $\vvv$ and $\hat{\sigma_i}$, the unit vector from the
center of the scatterer to the point of incidence at collision $i$, as
$v\cos\theta_i =|\vvv\cdot\hat{\sigma_i}|$, with the incident velocity
and unit vector oriented so that
$\vvv\cdot\hat{\sigma_i}\leq 0$.. The time between one collision of the moving
particle and the next is sampled from the normalized equilibrium
distribution of free times,\, $p(\tau)$,\, given for low densities, by
\be
p(\tau) = \n e^{-\n\tau},
\label{5}
\ee
where $\n$ is the low density value of the average collision
frequency, given for $d$-dimensional dilute, random Lorentz gases by 
\be
\nu = 2nv a^{d-1}\frac{\pi^{\frac{d-1}{2}}}{
(d-1)\Gamma(\frac{d-1}{2})}.
\label{5a}
\ee
Using this expression, we can write the total stretching
factor for a sequence of $\cn$ collisions of the moving particle as
\be
\Lambda(\vvr,\vvv,\cn) =\prod_1^\cn \Lambda_i^{(d)}(\tau_i,\theta_i).
\label{6}
\ee
The total time, $T$ between the
initial instant, and the final collision is $T=\tau_1 +\ldots +\tau_{\cn}$.

For low densities we can provide an expression for the average value
of the stretching factor for a sequence of ${\cn}$ uncorrelated
collisions 
taking place within a time interval $t$ where the first collision
takes place at time $\tau_1$ and the last at time $T$. This
average includes the probability that ${\cn}$ collisions will take
place within a time $t$, so these averages will directly determine the
dynamical partition function. The average
value of the stretching factor is given by
\bea
&&\left<[\Lambda
(\vvr,\vvv,\cn)]^{(1-\ba)}\right>_t  = 
\n(\frac{\n}{J_d})^{\cn}\int_0^{\infty}d\tau_1
\cdots \int_0^{\infty}d\tau_{\cn +1}\int^{\prime}\,d{\hat{\sigma}}_1,
\ldots d{\hat{\sigma}}_{\cn}
\prod_1^{\cn}\cos\theta_i  \times  \nonumber
\\
&&
\left[\prod_{i=1}^{\cn}\Lambda_{i}^{(d)}(\tau_i,\theta_i)\right]^{(1-\ba)}
e^{-\nu(\tau_1+\cdots +\tau_{\cn}+\tau_{\cn +1})}
\left[\Theta(t-\sum_1^{\cn}\tau_i)-\Theta(t-\sum_1^{\cn+1}\tau_i)\right].
\label{7}
\eea
Here $\Theta(x)$ is the usual Heaviside function. The Heaviside
functions and the additional time integral, over $\tau_{\cn +1}$,
are included in this expression to require that precisely ${\cn}$ collisions
take place over time $t$. The averaging
includes averages over all possible free times between two collisions,
using Eq. (\ref{5}), as well as integrations over the possible
directions of incidence at each collision. The prime on the
integrations over solid angles indicates that only half of the total
solid angle is to be included corresponding to the requirement that
$\vvv\cdot\hat{\sigma}\leq 0$, at each collision. The 
angular factors of $\cos\theta_i$ 
properly account for the volumes of collision cylinders when one
calculates the rate at which collisions take place with angle of
incidence, $\theta_i$, and $J_d$ is a normalization constant used in
the averaging and is given, for $d$-dimensions, by
\be
J_d =\frac{2\pi^{\frac{d-1}{2}}}{(d-1)\Gamma(\frac{d-1}{2})}. 
\ee
The dynamical partition function for the Lorentz gas at low density can be 
expressed in 
terms of these
average values by summing over contributions from all uncorrelated collision sequences
as
\be
Z(\ba,t)
=\sum_{\cn=0}^{\infty}\left<[\Lambda(\vvr,\vvv,{\cn})]^{1-\ba}\right>_t.
\label{8}
\ee
The term for ${\cn}=0$, will be set equal to $\n\exp[-\n t]$,
corresponding to the dynamical partition function for a particle with
no collisions in the time interval $(0,t)$ and a stretching factor of unity.

\section{The Zeta Function}

Although the integrals in each term in Eq. (\ref{8}) are not
difficult, the calculation of the topological pressure is greatly
simplified by introducing a zeta function \cite{ruelle},  adapted
to continuous time dynamics, obtained by taking the Laplace transform on Eq. (\ref{8}). One finds, 
using 
Eq. (4), that
\bea
{\cal{Z}}^{(
d)}(z) & = & \int_0^{\infty} dt e^{-zt}\,Z(\ba,t) = \\ \nonumber
& = & \frac{1}{
\nu+z}\left\{1-
G(d,\ba)\frac{d-1}{2} \frac{\n}{(z+\n)^{(d+\ba -d\ba)}}\left(\frac{2v}{a}\right)^{(d-1)(1-\ba)}\right\}^{-1}
\label{9}
\eea
where
\be
G(d,\ba)=\frac{\Gamma(\frac{d-1}{2})\Gamma(\frac{d-1
      +\ba(3-d)}{2})\Gamma(d+\ba-d\ba)}{\Gamma(d-1 +\frac{\ba(3-d)}{2})}.
\label{9a}
\ee

The pole
of the zeta function which is closest to the origin is the topological pressure. 
This 
follows simply 
from the observation that if the dynamical partition behaves, for asymptotically 
large 
$t$ as $\exp[tP(\ba)]$, the zeta function will have a pole at $z=P(\ba)$. It is 
an 
elementary calculation to find this pole, which is given by 
\be
P^{(d)}
_{mf}(\ba) = 
\left[\n\frac{d-1}{2}(\frac{2v}{a})^{(d-1)(1-\ba)}G(d,\ba)\right]^{\frac{1}{d+\ba-d\ba}} - \n.
\label{11}
\ee
Here the subscript $_{mf}$ indicates
that these are ``mean field'' results,
obtained by ignoring correlations between successive collisions. As we
shall see in the next sections, these results are valid only in a
small region about $\beta =1$
for large systems.

We can now use these results for the pressure to check if they agree
with known results and to determine the values for other dynamical
quantities. The 
topological pressure for $\ba = 1$,
should vanish for a closed system, and one easily sees from Eq.
(\ref{11}), using simple identities for gamma functions, that this expression satisfies this
condition. Further, the KS entropy can be determined from the
topological pressure by taking a derivative with respect to $\ba$ and
setting $\ba=1$, as
\be
h_{KS} = -\left. \frac{dP(\ba)}{d\ba}\right|_{\ba=1}
\label{13}
\ee
Simple calculations show that this condition is satisfied as well,
leading,
for $d=2,3$ to the results found by 
Van Beijeren, Latz, and Dorfman~\cite{2d,3d}:
\bea
h^{(2)}_{KS} = 2nav(1 - \gamma -\ln(2na^{2})); \\
\label{h2}
h^{(3)}_{KS} = 2na^{2}v \pi(\ln 2-\gamma -\ln(na^3\pi)).
\eea
Here $\gamma$ is Euler's constant.

New results can be obtained from the pressure by setting $\ba =0$ . The
value of the pressure at this point is the topological entropy per
unit time, and we thus obtain the mean field value
\be
h^{(d)}_{top} = \left[nv^d(d-1)(4\pi)^{\frac{d-1}{2}}\Gamma(\frac{d-1}{2})\right]^{\frac{1}{d}}-\n,
\label{14}
\ee
where we have inserted expression Eq. (\ref{5a}) for the collision
frequency in the first term on the right hand side of Eq. (\ref{14}).
It is interesting to note that these topological entropies depend upon
the $1/d$-th power of the density of the scatterers, and have
a finite, non-zero limit as the size of the scatterers vanishes. This
latter result is consistent with rigorous results of 
Burago,
Ferleger, and Kononenko\cite{burr} who provided estimates for the
topological entropy of the
periodic Sinai billiard in $d$-dimensions. They were
able to prove that for this system, the topological entropy has a
finite non-zero limit as the radius of the scatterers
vanishes, and
that in this limit, 
when variables are used in which both the density and the velocity equal unity,
the topological pressure is a non-decreasing
function of the number of dimensions,
bounded from below by $\ln(2d-1)$.
 One easily checks that these properties are also 
satisfied by the disordered Lorentz gas. For large $d$ in fact one finds an increase
proportional to $\sqrt{d}$ rather than $\ln d$.
However, please note that with increasing system size,
for fixed radius $a$, the present expressions 
for the topological pressure
of the disordered Lorentz gas, for $\beta$-values well below unity, soon have to 
be replaced by values resulting from orbits restricted to regions with high 
scatterer density.
Therefore the validity of Eq.~(\ref{14}) is 
restricted to a maximal system size, depending in turn on the density of   
scatterers. We will come back to this in our discussion.

\section{Large Systems}\label{sec:largesys}
For chaotic systems the dynamical instability typically gives rise to an   
exponential increase with time of the stretching factor $\Lambda(\vvr,\vvv,t)$.
For a given time, the actual rate of increase will depend on the initial values of 
$\vvr$ and $\vvv$. For $\ba>1$ regions of slow increase will be weighted most 
heavily. Therefore such regions may dominate the dynamical partition function, 
even if the probability of the moving particle to stay inside it decays 
exponentially with time. Our claim is that for specific realizations of the 
disordered Lorentz gas the orbits dominating the dynamical partition function
for long times at $\ba>1$ are orbits concentrated around the least unstable 
periodic orbit 
(LUPO) of the system, that is, the periodic orbit with the smallest exponential 
increase of its stretching factor. We have no rigorous proof for this statement, 
but we think it is extremely plausible. And in any case, the resulting values 
for the topological pressure do establish strict lower bounds to the actual 
values.

For $\ba<1$ the dynamical partition function will tend to be dominated by orbits 
in regions with a higher than average stretching factor. In analogy with the 
case $\ba>1$, one might expect that, for $\ba <1$, the dominating region will consist of orbits 
concentrated around the most unstable periodic orbit
\footnote{We
  observe from Eq. (\ref{4}) that the stretching factor of an orbit may
  be large for $d=2$, and for grazing collisions where $\theta \approx
  \pi/2$. However for a disordered Lorentz gas it is not possible to
find long orbits mainly consisting of nearly grazing collisions, and
the typical contribution to the topological pressure for $\ba <1$ will
come from orbits of the type discussed here.}. For $\ba<0$ this is indeed 
the case, for large enough systems, but for $\ba$ between 0 and 1 this is not 
true. The reason is that this set of orbits picks up a weight factor 
$1/\Lambda(\vvr,\vvv,t)$ for its long term survival probability, hence one sees 
from Eq.\ (\ref{1}) that for $\ba$ between 0 and 1 periodic orbits with a large 
stretching factor get small weight in the dynamical partition function.
Instead one has to look for the spatial region with the optimal combination of 
low escape rate and high average stretching factor for orbits confined to it.
Here one recognizes a strong analogy to the Donsker-Varadhan arguments~\cite{dv} 
for stretched exponential decay in diffusion among fixed traps.

In the remainder of this section we separately discuss the cases $\ba>1$ and 
$\ba<1$ 
in greater detail.

\subsection{The topological pressure for $\ba>1$\label{subsec:ba>1}}
\begin{figure}[h]
\centerline{\epsfig{file=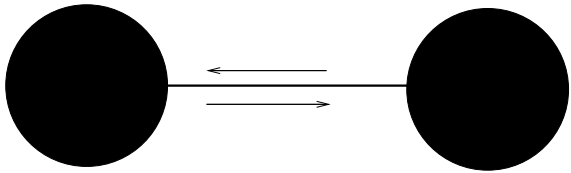,height=4cm}}
\caption{A typical unstable periodic orbit. The light particle bounces back and 
forth between the same two scatterers.}
\label{fig:period2}
\end{figure}

The least unstable periodic orbit is, roughly speaking, the orbit with the 
smallest 
number 
of collisions per unit time. In almost all cases this will simply be an orbit 
with the light particle bouncing back and forth between the same two scatterers, 
as illustrated in Fig.~\ref{fig:period2}. In exceptional cases though, the LUPO 
may involve more than two scatterers.

Assuming that indeed the LUPO involves just two scatterers and, for simplicity,
that the density of scatterers is low, we easily find an approximation for the 
distribution of the length of the free, periodic path between the two 
scatterers. Let $P(l)$ denote the probability that the system nowhere contains a 
periodic path of the type sketched above with a free path length exceeding $l$. 
For 
a large system, with $N$ scatterers in a volume $V$, this probability, in two 
dimensions, satisfies an equation which is approximately
\be
\frac {d\,P(l)}{d\,l}=\pi n(l+2a)N e^{-2nal}P(l),
\label{pl}
\ee
which is obtained by the following reasoning: suppose the system has no periodic 
paths with a free path length exceeding $l$. The chance for this is $P(l)$. To 
obtain the probability 
for a largest periodic path between $l$ and $l-d\,l$ one has to multiply this 
with the probability of having two scatterers at a distance between $l+2a$ and 
$l+2a-d\,l$, which is roughly $N^2 2\pi(l+2a)d\,l/(2V)$, times the probability, 
$\exp\,(-2nal)$, that the periodic orbit between those scatterers is free from 
other scatterers. Integrating this equation and taking the derivative of $P(l)$ 
with respect to $l$, one finds the probability distribution $p(l)$ for the 
length of the LUPO as
\be
p(l)=\pi
n N(l+2a) e^{-2nal}\exp\left\{-\frac{\pi N}{2a}\left(l+2a+\frac 
1{2na}\right) e^{-2nal}\right\}.
\label{lupo}
\ee
For large $N$ this distribution is sharply peaked around a value $l_0$ 
satisfying
\be
l_0=\frac{\log\frac{\pi N(l_0+2a)}{2a}}{2na}\ \ \ \ \ d=2.
\label{l0}
\ee
The width of the distribution is of $O(1)$, and therefore of $O(1/\log N)$ 
relative to $l_0$.
In three dimensions a completely analogous calculation leads to
\be
P(l)=\exp\left\{-2N\left(\frac{(l+2a)^2}{a^2} +\frac{2(l+a)}{\pi na^4} +\frac 
2{\pi^2n^2a^6}\right)e^{-\pi na^2l}\right\}.
\label{pl3}
\ee
From this one finds the maximum $l_0$ in the distribution of the length of the 
LUPO satisfies the equation
\be
l_0=\frac{\log\frac{2(l_0+2a)^2N}{a^2}}{\pi na^2}\ \ \ \ \ d=3.
\label{lupo3}
\ee
Again, one finds that 
the distribution is sharply peaked around a value proportional to $\log 
N$, with fluctuations of $O(1)$.
Here too the results could easily be extended to arbitrary dimensionality $d$,
but we did not work this out.

For a given scatterer configuration the topological pressure resulting from the 
orbits near the LUPO follows from the value of $l_0$, as
\be
P^{(d)}_{po}(\ba)=-\ba\frac{(d-1)v}{l_0}\log\frac{2l_0}{a},
\label{plupo}
\ee
where we used Eq.\ (\ref{4}), with $\tau_i=l_0/v$ and $\phi_i=0$, and the fact 
that the long time survival probability near an unstable periodic orbit of 
stretching factor $\Lambda(t)$ is given by $1/\Lambda(t)$.

For given $\ba$ and $l_0$ the topological pressure is found as the larger of
the expressions (\ref{11}) 
and (\ref{plupo}). Since the 
former are proportional to $\ba-1$ and the latter roughly to $\ba/\log N$, there 
is a sharp phase transition at a $\ba$-value satisfying $\ba-1 \sim 1/\log N$. 
In 
the terminology of ordinary thermodynamics this is a {\em first order phase 
transition}, as the first derivative of the topological pressure with respect to 
$\ba$ is discontinuous at this transition. This implies, however, that the range 
of $\ba$-values $>1$ for which the topological pressure gives information about 
the bulk properties of the system is very limited for large systems; 
specifically 
it is of order $1/\log N$. For larger $\ba$-values the topological pressure 
rather gives information about the smallest escape rate of orbits from the 
neighborhoods of unstable periodic orbits. This obviously is of great importance 
for the asymptotic rate of decay to equilibrium, which cannot be larger than 
this 
smallest escape rate. It can easily be smaller, if the asymptotic rate of mixing 
is smaller than the smallest escape rate or, even worse, if the system does not 
decay to equilibrium at all. Notice that the smallest escape rate may be zero, 
corresponding either to algebraic escape or to finite probability of not 
escaping 
at all (in which case one should not speak of {\em unstable} periodic orbits any 
more). But one should also notice that the pockets in phase space corresponding 
to the neighborhoods of unstable periodic orbits often cover such minute  
fractions of the total available phase space volume, that for the physics of the 
system they are of no practical importance.
We will come back to this briefly in our discussion.

\subsection{The topological pressure for $0<\ba<1$\label{subsec:ba<1}}
\begin{figure}[h]
\centerline{\epsfig{file=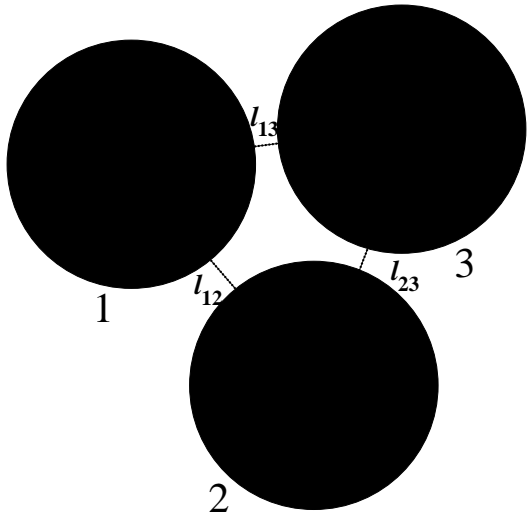,height=8cm}}
\caption{A region enclosed by three disks. The escape rate of a light particle 
from this region is small and the collision frequency with the surrounding 
scatterers is high. The total escape length $l$ is the sum of $l_{12},l_{13}$ 
and 
$l_{23}$.}
\label{fig:3disks}
\end{figure}

For $0<\ba<1$ the factor $1/\Lambda(t)$ in the weighting of the 
near periodic orbits suppresses the contributions of strongly 
diverging periodic orbits to the topological pressure, as we 
noted already. Instead, we have to look for compact regions with 
high collision rate and slow escape. In two dimensions, with scatterers not 
allowed to overlap each other, the regions best satisfying these criteria 
obviously are enclosures of three scatterers almost touching each other, as 
illustrated in Fig.~\ref{fig:3disks}. Again, we have no rigorous proof of this, 
but the statement seems even more obvious than the one about the LUPO's.  

The escape rate from such a trapping region may be expressed to a very good 
approximation as\cite{maczw}
\be
\nu_{tr}=\frac{vl_{esc}}{\pi O},
\label{nutr}
\ee
with $O\approx (\sqrt{3}-\half \pi)a^2$ the surface area of the trapping region 
and the escape length $l_{esc}$, the sum of the distances between the pairs of 
surrounding spheres.

In  large systems there will be many of such trapping regions and the
one with the smallest $l_{esc}$ will determine the topological
pressure when $\ba$ is sufficiently smaller than $1$. 
Let us call this value $l_1$. We define $Q(l)$ as the probability that none of 
the trapping regions has an escape length $<l$ and notice that it approximately 
satisfies the equation
\be
\frac{d\,Q(l)}{d\,l}=- \frac 1 6 \sqrt 3 \pi a l^2 n^2 N \chi_3 Q(l),
\label{ql}
\ee
with $\chi_3$ the triplet correlation function for three scatterers, all 
touching 
each other.
The solution to this equation is
\be
Q(l)=\exp\left(-\frac{ \sqrt 3 \pi a l^3 n^2 N\chi_3}{18}\right),
\label{solQ}
\ee
from which the distribution of $l_1$ is obtained as
$q(l_1)=-d\,Q(l_1)/d\, l_1$. The maximum of this distribution is not
sharp, but for our purpose, it is only important  that this maximum scales as a 
function of $l_1/(N^{1/3}a)$. Consequently, as a function of $N$ the escape rate 
$\nu_{tr}$ scales in the same way, even though its specific value depends on the 
scatterer configuration at hand.

The topological pressure in this case may be expressed as
\be
P^{(2)}_{tr}(\ba)=(1-\ba)\lambda^{SB}_{cp}-\nu_{tr},
\label{ptr}
\ee
with $\lambda^{SB}_{cp}$ the positive Lyapunov exponent of a triangular Sinai 
billiard at close packing. From Ref.~\cite{SBvalue} we quote the value
$\lambda^{SB}_{cp}\approx 3.6v/a$.
Comparing (\ref{ptr}) and (\ref{11}) we note that the range of
$\ba$-values over which the mean-field value usually dominates that of
the most trapping region satisfies
\be
\ba > 1-\frac{\nu_{tr}}{\lambda_{cp}^{SB}-h_{KS}^{(2)}},
\ee
where $h_{KS}^{(2)}$ is given by Eq. (\ref{h2}). Thus, the validity of the
mean field result
is restricted in this case to a region of $\ba$ values slightly below
unity, with size of  $O(N^{-1/3})$.
Here, too, the phase transition that occurs when the most trapping region 
becomes 
dominant will be of first order.

The three dimensional case
(again, for simplicity we won't consider arbitrary $d$) is more subtle because, with
non-overlapping scatterers, there are no trapping regions from which
escape only is possible through very narrow channels. Instead, a large
topological pressure will result from the presence of a fairly large
compact volume with a higher than average scatterer density. Thus,
such high density regions will determine the topological pressure,
even in three dimensions, away from a small region near $\ba=1$. In
order to obtain useful estimates of the size of this region, we
consider compact volumes of radius $R$, typically of spherical shape,
though this particular shape is not essential to our argument.  The
escape rate from such a region will be of the form $\nu(R)\sim D/R^2$,
with $D$ the diffusion coefficient. Here we use the fact that the
distribution of moving particles in the disordered Lorentz gas
satisfies a diffusion equation on large time and length scales. The
largest density fluctuation found in such a region in a system of
total volume $R^3$
 will typically be $\sim n^{1/2}R^{-3/2} \log^{1/2}
(N/nR^3)$.
This follows from the Gaussian nature of density fluctuations in large regions, 
with standard deviation $\sim n/(R^3)$, and the observation that the number of 
independent volumes of radius $R$ is proportional to $N/nR^3$.
Therefore the excess topological pressure resulting from
trajectories restricted to a region of radius $R$ with maximal density
fluctuation is obtained by means of a simple Taylor expansion of the
topological pressure about its mean field value in
powers of the density
deviation. The resulting correction to the mean field
pressure will then be roughly of the form
\be
\Delta P^{(3)}_{tr}(\ba)=C_1(1-\ba)\frac{n^{1/2}}{R^{3/2}}\frac{\partial 
c^{(3)}_{mf}}{\partial n}\left(\log\frac{N}{nR^3}\right)^{1/2}-\frac{C_2 
D}{R^2},
\label{DeltaP}
\ee
with $C_1$ and $C_2$ constants of order unity and
the subscript $_{tr}$ indicating that these corrections are due to trapping 
regions.
Furthermore,
\[
c^{(3)}_{mf}\equiv \frac{P^{(3)}_{mf}(\ba)}{1-\ba}.
\]
The expression (\ref{DeltaP}) takes its largest value, $\sim(1-\ba)^4(\log 
N)^2$, 
for $R$ roughly proportional to 
$(1-\ba)^{-2}(\log N)^{-1}$. In order for this largest value of
$\Delta P^{(3)}_{tr}$ to be a small
correction to $P^{(3)}_{mf}(\ba)$, the parameter $\ba$ must satisfy
the condition $1-\ba\ll (\log N)^{-2/3}$. Furthermore, for $\Delta 
P^{(3)}_{tr}(\ba)$ to be negative for all allowable  values of $R$ one needs
$1-\ba<C N^{-1/6}$, with $C$ some constant.

On the basis of this analysis one can expect that also the phase
transition from $P^{(3)}_{mf}$ to $P^{(3)}_{tr}$ will be first order,
but only very, very weakly so. If $\ba$ is increased within the range
$1-\ba\sim N^{-1/6}$ one finds that for $\ba$ very close to $1$, 
the expression (\ref{DeltaP}) is negative definite within the
allowable range of $R$-values. It first becomes positive, at a value
of $R/V^{1/3}$ where the expression on the right hand side of
Eq. (\ref{DeltaP}) attains a maximum, for some specific $\ba$, which
then marks the phase transition. Since at this transition $R$ jumps to
a value $\ll V^{1/3}$ it is a first order transition indeed.
However, the resulting change in the topological pressure is so
small, for $\ba$ in the range between $1-\ba\sim N^{-1/6}$ and
$1-\ba\sim (\log N)^{-2/3}$, that in fact the transition will appear
to be continuous. Furthermore one should keep in mind that the scenario sketched 
here for the determination of the topological pressure is very heuristic. There 
may be additional contributions to the dynamical partition function, which we 
have overlooked, but are in fact dominant for certain ranges of $\ba$-values.

Further, we want to remark that Eq. (\ref{DeltaP}) should be used with
care for $1-\ba$ strongly exceeding $(\log N)^{-2/3}$. For the
diffusion approximation for the escape rate to be valid $R$ should be
much larger than both the size of the scatterers, $a$,  and the mean
free path between collisions. Further, for the Taylor expansion of the
topological pressure to be valid,
the expression multiplying ${\partial c^{(3)}_{mf}}/{\partial n}$ on
the right hand side of Eq. (\ref{DeltaP}) should be $\ll 1$ . Nevertheless, in 
this range of $\ba$-values we can be sure that for large systems the topological 
pressure is dominated by contributions from one of the trapping regions.

It is also useful to note that even for $\ba>1$ there is a small
region that is dominated by trajectories trapped in compact regions,
this time in regions of {\em lower than average} density. Combining the 
arguments 
presented before, one finds that this region is determined roughly by conditions 
of the form $C_3 N^{-1/6}<\ba-1<C_4(\log N)^{-3/4}$, with $C_3$ and $C_4$  
constants of order unity. However, throughout this region the relative 
corrections to $P^{(3)}_{mf}$ are small.

\section{Discussion}
In summary, we have obtained explicit, mean-field type expressions for the 
topological, or Ruelle pressure of a dilute disordered Lorentz gas, leading to  
Kolmogorov-Sinai entropies in agreement with the results of previous 
calculations. 
However, we also found that for large systems the range of the 
inverse-temperature like parameter $\ba$ of Ruelle's thermodynamic formalism 
over 
which these expressions
are valid, is restricted to a very small region around $\ba=1$. For
$\beta>1+O(1/\log N)$, with $N$ the number of scatterers in the
system, the topological pressure is determined by the decay rate of
orbits from the region in phase space surrounding the least unstable
periodic orbit, and for $\ba<1-O(N^{-1/3})$ in two dimensions, and for
$\ba<1-O((\log N)^{-2/3})$ in three dimensions it is determined by
orbits that remain trapped in areas enclosed by three disks, or in
fairly large regions of higher than average scatterer density, respectively. In 
fact these restrictions do not only apply to dilute systems. They will hold for 
any system with particles moving among a disordered array of fixed scatterers.

The periodic orbit results are of interest because they give information related 
to the asymptotic long time decay to equilibrium. Similarly the trapped region 
results
contain information about the fastest rate of information loss that may be 
observed for long times, but the way in which different regions are weighted, 
especially how their escape rates are taken into account, depends on the choice 
of $\ba$. 

Another context where the least unstable periodic orbits play a crucial role is 
in the description of {\em scars}, i.e.\ eigenfunctions in quantum systems that 
are concentrated in the neighborhoods of periodic orbits of the corresponding 
classical system~\cite{scars}. An important difference is that in this case the 
classical orbit need not have a minimal length (or more precisely, a stretching 
factor that is sufficiently smaller than the average bulk stretching factor) to 
be observable as a scar. Nevertheless, our findings suggest that the topological 
pressure for $\ba>1$ may be used as a tool for finding the most prominent scar
in the quantum mechanical counterpart of a classically chaotic system.

Two of the most interesting questions to be asked are: In how far
can the
results obtained here
be generalized to systems of interacting
moving particles? and: 
In how far will similar conclusions
 hold in such
cases? For simple gases at low densities we believe that we can
calculate the leading order mean field approximation to the
topological pressure along the lines sketched in the present paper, by
combining these with the methods of \cite{kshs}. Again, for $\ba<1$ we
expect dominance of near periodic orbits. In fact, for hard spheres
the situation is even slightly worse than in the Lorentz gas. With
periodic or simple reflecting boundary conditions one may easily
construct periodic orbits without any collisions, simply by assigning
equal velocities to all particles (or, if one insists on zero total
momentum, velocity $\vvv$ to half of the particles, in one half of the
box, and $-\vvv$ to the other half, in the other half of the
volume). The neighborhoods of these orbits will give rise to a strictly
vanishing topological pressure for all $\ba>1$. Even with irregularly
shaped boundaries and not too many particles, one may construct orbits
without inter-particle collisions by lining up all the particles in a
long row, or ``duck march'', along the same single particle trajectory
in phase space, chosen such that there are no near-intersections with
a short time lag. There may be a stretching factor $>1$ for such an
orbit, but that is entirely due to the collisions with the boundaries
and therefore will be anomalously small compared to the stretching
factors for typical initial conditions. One must say that it looks 
unsatisfactory 
when the topological pressure
of a system of many interacting particles is determined exclusively by
a very rare process in which the only collisions occurring are with the 
boundaries. This obviously is a case where the long time asymptotics of the 
equations of motion in tangent space is completely irrelevant from the physics 
point of view.
It is a clear example illustrating once more why the large system limit is both 
so hard and so interesting.

For $\ba<1$ our conclusion that the topological pressure soon becomes dominated 
by orbits constrained to a denser than average part of space, does not
carry over in this way to systems of many interacting particles. There
are no stationary regions in such systems that will maintain a higher
than average density forever. On the other hand, this phenomenon is a
strong warning that similar behavior might very well occur in fluid
systems. For example, if the phase space of a many particle system contains 
metastable regions with local stretching factors that are larger than the 
average 
stretching factor of a ``typical" equilibrium 
region, the topological pressure for $\ba<1-\epsilon$ most likely will
be determined by orbits constrained to one of these metastable
regions. In such a case again, the topological pressure contains no information 
on the equilibrium properties of the system, but it is a fascinating idea that a 
study of the $\beta$-dependence of the topological pressure may reveal 
properties 
of the metastable states of a system.
A concrete example where such dominance may be expected, in our
opinion is provided by hard spheres at densities beyond
solidification. For such densities there are many glassy states, as is
known for example from experiments on colloidal
systems~\cite{coll}. It is very likely that the collision frequency in
these glassy states is higher than in the crystalline state, and the
corresponding stretching factor will be higher.

We also remark that it may be very helpful to use the Ruelle
zeta-function not only for extracting the topological pressure from
its pole, but also for studying its full $z$-dependence in more
detail. For small, but not too small, $z$ one might expect this
function to exhibit the behavior of ``typical stretching factors" in
the bulk system. A possibility would be to examine various scalings of
$z$ with system size and to consider the behavior of the zeta-function
in terms of its dependence upon the scaled parameter in the infinite system 
limit, much in the spirit of hydrodynamic scaling.

We also want to add some words of caution regarding the topological entropy, in 
other words the topological pressure for $\ba=0$. 
For not even very large systems this value of $\ba$ is far beyond the
region in which the topological pressure is determined by bulk
averages. Therefore, in disordered systems, the topological entropy,
rather than revealing how the explored area in phase space of a
typical trajectory bundle increases with time, will merely provide information, 
in most cases, about very atypical trajectory bundles, restricted to some small 
subspace of phase space.

Finally, we plan to extend the work reported here, not only to dilute gases of 
moving particles, but also to Lorentz gas systems with open boundaries and/or 
with driving fields combined with Gaussian thermostats.

{\acknowledgments}

We are very pleased and honored to dedicate this paper to David Ruelle
and Yasha Sinai, whose fundamental and fruitful contributions to the fields of 
dynamical systems theory and statistical mechanics, we greatly admire and 
heavily 
use. The authors thank the referee for calling their attention to the
work of Burago, Ferleger, and Kononenko on the topological pressure of
billiard systems, and for other helpful remarks.
HvB acknowledges support by the National Fund for Scientific Research
(F.\ N.\ R.\ S.\ Belgium) and by  the Mathematical Physics program of
FOM and NWO/GBE. JRD acknowledges support from the National Science
Foundation (USA) under grant NSF-PHY-98-20824.

\end{document}